\documentclass{svjour3}                     % onecolumn (standard format)
\smartqed  % flush right qed marks, e.g. at end of proof
\usepackage{graphicx}
\usepackage{fix-cm}
\usepackage{amsmath}
\journalname{International Journal of Theoretical Physics}

\begin{document}

\title{Conformal symmetries of spherical spacetimes}


\author{S. Moopanar        \and
        S. D. Maharaj}


\institute{
S. Moopanar        \and S. D. Maharaj \at
Astrophysics and Cosmology Research Unit,
 School of Mathematical Sciences, Private Bag X54001,
 University of KwaZulu-Natal, Durban 4000, South Africa
\\
\email{maharaj@ukzn.ac.za} }

\date{Received: date / Accepted: date}

\maketitle

\begin{abstract}
We investigate the conformal geometry of spherically symmetric
spacetimes in general without specifying the form of the matter
distribution. The general conformal Killing symmetry is obtained
subject to a number of integrability conditions. Previous results
relating to static spacetimes are shown to be a special case of our
solution. The general inheriting conformal symmetry vector, which
maps fluid flow lines conformally onto fluid flow lines, is
generated and the integrability conditions are shown to be
satisfied. We show that there exists a hypersurface orthogonal
conformal Killing vector in an exact solution of Einstein's
equations for a relativistic fluid which is expanding, accelerating
and shearing.


\end{abstract}

\section{Introduction} \label{sec:1}

Conformal symmetries are studied in general relativity because they
preserve the structure of the null cone. Together with conventional
isometries, which have been more widely studied, they help to
provide a deeper insight into the spacetime geometry. They assist in
the generation of exact solutions, sometimes new solutions, of the
Einstein field equations which may be utilised to model
relativistic, astrophysical and cosmological phenomena.
 Stephani \textit{et al}.~\cite{stephani} have emphasized the role of symmetries in
classifying and categorising exact solutions. Symmetries are used as
one of the principal classification schemes in their catalogue of
known solutions. However there are few conformal symmetries known
even in spacetimes of high symmetry. Clearly there is a need to
study conformal symmetries in a general relativistic setting.

In recent treatments conformal symmetries have been studied by Keane
and Barrett~\cite{keane1} in Robertson--Walker spacetimes, Tupper
\textit{et al}.~\cite{tupper} in null spacetimes and Keane and Tupper
\cite{keane2}  in \textit{pp}--wave spacetimes. We have chosen to
investigate the existence of conformal symmetries in inhomogeneous
spherically symmetric spacetimes because of their physical
significance. Our intention is to generate the general conformal
Killing vector without making any additional assumptions.

In Sect.~\ref{sec:2}  the line element and the conformal Killing
vector equations are presented; the conformal equations are solved
in general and the solution can be compactly written subject to
integrability conditions. The results of Maharaj \textit{et al}.~\cite{maharaj2}
 are regained for the special case of static spherically
symmetric spacetimes in Sect.~\ref{sec:3}. We find the general
inheriting conformal symmetry vector, which maps fluid lines
conformally, by fully integrating the integrability conditions in
Sect.~\ref{sec:4}. We demonstrate, in Sect.~\ref{sec:5}, that an
exact solution of the Einstein field equations for an expanding,
shearing and accelerating relativistic fluid admits a hypersurface
orthogonal conformal Killing vector.

\section{Conformal equations} \label{sec:2}

The line element for the general spherically symmetric spacetimes
can be written as
    \begin{equation}
    ds^2 = -e^{2\nu(t,r)}dt^2 + e^{2\lambda(t,r)} dr^2
    + Y^2(t,r) (d\theta^2 +\sin^{2}\theta \,d\phi^{2}) \label{metric}
    \end{equation}
where the functions $\nu$, $\lambda$ and $Y$ represent the
gravitational potentials. In the representation (\ref{metric}) we
have used comoving coordinates $(x^a) = (t,r,\theta,\phi)$ relative
to the fluid 4--velocity $\displaystyle u^a = e^{-\nu}
\delta^a_{{}0}$. By spherical symmetry the vorticity has to vanish
in comoving coordinates. However the acceleration, expansion and the
shear are the nonvanishing kinematical quantities in general. In the
special case of spherically symmetric spacetimes which are static we
can adapt the comoving coordinates to write the line element in the
Schwarzschild form
\begin{equation}
ds^2 = -e^{2\nu(r)}dt^2 + e^{2\lambda(r)} dr^2
    + r^2  (d\theta^2 +\sin^{2}\theta \,d\phi^{2}). \label{static}
\end{equation}
The metric (\ref{metric}) is utilised in constructing inhomogeneous
cosmological models (see Stephani \textit{et al}.~ \cite{stephani}) and the metric
(\ref{static}) is used in describing dense stellar astrophysical
models (see Delgaty and Lake \cite{delgaty}).

A conformal Killing vector $\bf X$ is defined by the action of the
Lie infinitesimal operator ${\cal L}_{\bf X}$ on the metric tensor
field $\bf g$ so that
\begin{equation}
{\cal L}_{\bf X} g_{ab} = 2 \psi g_{ab}      \label{lie}
\end{equation}
where $\psi = \psi(x^a)$ is the conformal factor. In this section we
analyse the full conformal symmetry for spherically symmetric
manifolds without making any assumptions about the conformal vector
$\bf X$ or the conformal factor $\psi$.

The conformal equation (\ref{lie}) for the metric (\ref{metric})
reduces to the following set of equations
\begin{subequations} \label{conf}
\begin{eqnarray}
\nu_t X^0 + \nu_r X^1 + X^0_{{}t} &=& \psi  \label{conf1} \\
e^{2\lambda} X^1_{{}t} - e^{2\nu} X^0_{{}r} &=& 0 \label{conf2}  \\
Y^2 X^2_{{}t} - e^{2\nu} X^0_{{}\theta} &=& 0 \label{conf3}  \\
Y^2 \sin^2 \theta  X^3_{{}t} - e^{2\nu} X^0_{{}\phi} &=& 0 \label{conf4} \\
\lambda_t X^0 + \lambda_r X^1 + X^1_{{}r} &=& \psi \label{conf5}  \\
Y^2 X^2_{{}r} + e^{2\lambda} X^1_{{}\theta} &=& 0 \label{conf6}  \\
Y^2 \sin^2 \theta X^3_{{}r} + e^{2\lambda} X^1_{{}\phi} &=& 0 \label{conf7}  \\
\frac{Y_t}{Y} X^0 + \frac{Y_r}{Y} X^1 + X^2_{{}\theta} &=& \psi  \label{conf8} \\
\sin^2 \theta X^3_{{}\theta} + X^2_{{}\phi} &=& 0 \label{conf9} \\
\frac{Y_t}{Y} X^0 + \frac{Y_r}{Y} X^1 + \cot \theta X^2 +
X^3_{{}\phi} &=& \psi \label{conf10}
\end{eqnarray}
\end{subequations}
which have been verified with the help of Mathematica. In the above
we have used the notation where subscripts denote partial
differentiation. The equations (\ref{conf}) comprise a coupled
system of partial differential equations which are first order and
linear in ${\bf X} = (X^0,X^1,X^2,X^3)$ and $\psi$.

The system (\ref{conf}) may be integrated in general to find the
components $X^0$, $X^1$, $X^2$, $X^3$ and $\psi$; this is possible
because we can manipulate the system to  obtain particular equations
involving only the individual components of the vector and the
conformal factor. In this process the equations decouple, and a
number of integrability conditions (on the metric functions $\nu$,
$\lambda$ and $Y$) are generated. It is a simple matter to check
that the given solution is valid by direct substitution in
(\ref{conf}). The components of the conformal Killing vector ${\bf
X}$ and the conformal factor $\psi$ are
\begin{eqnarray}
X^0 &=& Y^2 e^{-2\nu} \sin\theta ({\cal C}_t \sin\phi - {\cal D}_t
\cos\phi) - Y^2 e^{-2\nu} {\cal I}_t \cos\theta +
{\cal J}   \nonumber \\
X^1 &=& - Y^2 e^{-2\lambda} \sin\theta ({\cal C}_r \sin\phi - {\cal
D}_r \cos\phi) + Y^2 e^{-2\lambda} {\cal I}_r \cos\theta +
{\cal K}   \nonumber \\
X^2 &=& \cos\theta \left[ {\cal C} \sin\phi - {\cal D} \cos\phi
\right] + \cos\theta (a_1 \sin\phi - a_2 \cos\phi) - a_3 \sin\phi
\nonumber \\ & & + a_4 \cos\phi + {\cal I} \sin\theta \nonumber
\\
X^3 &=& \csc\theta \left[ {\cal C} \cos\phi + {\cal D} \sin\phi
\right] + \csc\theta (a_1 \cos\phi + a_2 \sin\phi) \nonumber \\ & &
- \cot\theta (a_3 \cos\phi
+ a_4 \sin\phi) + a_5 \nonumber \\
\psi &=& Y \sin\theta \sin\phi \left[Y e^{-2\nu}{\cal C}_{tt} +
(2Y_t - Y\nu_t ) e^{-2\nu}{\cal
C}_{t} - Y e^{-2\lambda} \nu_r {\cal C}_r \right] \nonumber \\
& & - Y \sin\theta \cos\phi \left[Y e^{-2\nu}{\cal D}_{tt} + (2Y_t -
Y\nu_t ) e^{-2\nu}{\cal D}_{t} - Y e^{-2\lambda} \nu_r {\cal D}_r
\right] \nonumber \\ & & - Y \cos\theta \left[ Y e^{-2\nu}{\cal
I}_{tt} + (2Y_t - Y\nu_t)
e^{-2\nu}{\cal I}_t - Y e^{-2\lambda} \nu_r {\cal I}_r \right] \nonumber \\
 & & + {\cal J}_t +
\nu_t {\cal J}  + \nu_r {\cal K} \nonumber
\end{eqnarray}
where ${\cal A}$, ${\cal C}$, ${\cal D}$, ${\cal I}$, ${\cal J}$,
${\cal K}$  are functions of $t$ and $r$ and $a_1$--$a_5$ are
constants. This general solution is subject to the following twelve
consistency conditions which place restrictions on the metric
functions:
\begin{eqnarray}
Y {\cal C}_{tr} + (Y_r - Y \nu_r) {\cal C}_{t}  +
(Y_t - Y \lambda_t) {\cal C}_{r} &=& 0   \nonumber \\
Y {\cal D}_{tr} + (Y_r - Y \nu_r) {\cal D}_{t}  +
(Y_t - Y \lambda_t) {\cal D}_{r} &=& 0   \nonumber \\
Y {\cal I}_{tr} + (Y_r-Y\nu_r) {\cal I}_{t} + (Y_t-Y\lambda_t) {\cal
I}_r &=& 0
\nonumber \\
Y e^{-2\nu}{\cal C}_{tt} + Y e^{-2\lambda}{\cal C}_{rr} +
(2Y_{t}-Y\lambda_t-Y\nu_t) e^{-2\nu}{\cal C}_{t} & & \nonumber \\
+ (2Y_r-Y\lambda_{r}-Y\nu_r) e^{-2\lambda} {\cal C}_{r}
&=& 0  \nonumber \\
Y e^{-2\nu}{\cal D}_{tt} + Y e^{-2\lambda}{\cal D}_{rr} +
(2Y_{t}-Y\lambda_t-Y\nu_t) e^{-2\nu}{\cal D}_{t} & & \nonumber \\
+ (2Y_r-Y\lambda_{r}-Y\nu_r) e^{-2\lambda} {\cal D}_{r}
&=& 0 \nonumber \\
Y e^{-2\nu} {\cal I}_{tt} + Y e^{-2\lambda}{\cal I}_{rr} +
(2Y_t - Y\lambda_t - Y\nu_t ) e^{-2\nu} {\cal I}_{t} & & \nonumber \\
+ (2Y_r - Y\lambda_r - Y\nu_r ) e^{-2\lambda} {\cal I}_{r}
&=& 0 \nonumber \\
Y^2 e^{-2\nu}{\cal C}_{tt} + Y (Y_t - Y\nu_t ) e^{-2\nu}{\cal C}_{t}
+ Y (Y_{r} - Y\nu_r ) e^{-2\lambda} {\cal C}_{r} + {\cal C} + a_1
&=& 0   \nonumber \\
Y^2 e^{-2\nu}{\cal D}_{tt} + Y (Y_t - Y\nu_t ) e^{-2\nu}{\cal D}_{t}
+ Y (Y_{r} - Y\nu_r ) e^{-2\lambda} {\cal D}_{r} + {\cal D} + a_2
&=& 0   \nonumber \\
Y^2 e^{-2\nu} {\cal I}_{tt} + Y (Y_t - Y\nu_t) e^{-2\nu} {\cal
I}_{t} + Y (Y_r - Y\nu_r ) e^{-2\lambda} {\cal I}_{r} + {\cal I}
&=& 0    \nonumber \\
e^{2\lambda} {\cal K}_{t} - e^{2\nu} {\cal J}_{r} &=& 0
\nonumber \\
- {\cal J}_t + \left( \frac{Y_t}{Y} - \nu_t \right) {\cal J} +
\left( \frac{Y_r}{Y} - \nu_r \right) {\cal K}
&=& 0    \nonumber \\
- {\cal J}_t + {\cal K}_r + (\lambda_t - \nu_t ) {\cal J} +
(\lambda_r - \nu_r ) {\cal K} &=& 0 \nonumber
\end{eqnarray}
Note that in our solution the angular dependence in $\theta$ and
$\phi$ is known explicitly. This feature of the solution is not
surprising as the spacetime is spherically symmetric. There is
freedom only in the $t$ and $r$ coordinates.

The above solution may be expressed more compactly if we utilise the
notation adopted by Maartens {\em et al}.~\cite{maartens3} in their
categorisation  of static spacetimes. We first let
\[
{\cal C} + a_1 = \tilde{\cal C} \qquad \mbox{and} \qquad {\cal D} +
a_2 = \tilde{\cal D}
\]
and introduce the new variables
\begin{eqnarray}
A^i &=& (A^1,A^2,A^3) = \left( \tilde{\cal C}(t,r), -\tilde{\cal
D}(t,r), -{\cal I}(t,r) \right)
\nonumber \\
\eta_i &=& (\eta_1,\eta_2,\eta_3) = (\sin\theta \sin\phi, \sin\theta
\cos\phi, \cos\theta)
\nonumber \\
A^0 &=&  {\cal J}(t,r) \nonumber \\
A^4 &=&  {\cal K}(t,r) \nonumber
\end{eqnarray}
The components of the conformal Killing vector and the conformal
factor then become
\begin{subequations} \label{soln}
\begin{eqnarray}
X^0 &=& Y^2 e^{-2\nu} A^i_{{}t} \eta_i + A^0 \label{soln1} \\
X^1 &=& -Y^2 e^{-2\lambda} A^i_{{}r} \eta_i + A^4 \label{soln2} \\
X^2 &=& A^i (\eta_i)_{{}_\theta} - a_3\sin\phi + a_4\cos\phi
\label{soln3} \\
X^3 &=& \csc^2\theta A^i (\eta_i)_{{}_\phi} - \cot\theta
(a_3\cos\phi+a_4\sin\phi) + a_6  \label{soln4} \\
\psi &=& Y \eta_i \left[ Ye^{-2\nu} A^i_{{}tt} + (2Y_t-Y\nu_t)
e^{-2\nu} A^i_{{}t} - Y e^{-2\lambda} \nu_r A^i_{{}r} \right] +
A^0_{{}t}
+ \nu_t A^0  \nonumber \\
& & + \nu_r A^4  \label{soln5}
\end{eqnarray}
\end{subequations}
The consistency conditions simplify to
\begin{subequations} \label{cons}
\begin{eqnarray}
Y A^i_{{}tr} + (Y_r-Y\nu_r) A^i_{{}t} + (Y_t-Y\lambda_t)
A^i_{{}r} &=& 0  \label{cons1} \\
Y e^{-2\nu} A^i_{{}tt} + Y e^{-2\lambda} A^i_{{}rr} +
(2Y_t-Y\lambda_t-Y\nu_t) e^{-2\nu} A^i_{{}t} \nonumber \\ +
(2Y_r-Y\lambda_r-Y\nu_r) e^{-2\lambda} A^i_{{}r} &=& 0
\label{cons2}
\\
Y^2 e^{-2\nu} A^i_{{}tt} + Y(Y_t-Y\nu_t) e^{-2\nu} A^i_{{}t} +
Y(Y_r-Y\nu_r) e^{-2\lambda} A^i_{{}r} + A^i &=& 0  \label{cons3}
\\
e^{2\lambda} A^4_{{}t} - e^{2\nu} A^0_{{}r} &=& 0  \label{cons4}
\\
-A^0_{{}t} + \left( \displaystyle\frac{Y_t}{Y} -\nu_t \right) A^0 +
\left( \displaystyle\frac{Y_r}{Y} -\nu_r \right) A^4 &=& 0
\label{cons5} \\
-A^0_{{}t} + \left( \lambda_t-\nu_t \right) A^0 + \left(
\lambda_r-\nu_r \right) A^4 + A^4_{{}r} &=& 0 \label{cons6}
\end{eqnarray}
\end{subequations}
It is remarkable that the general conformal symmetry (\ref{soln})
can be generated explicitly since the line element (\ref{metric})
represents the most general spherically symmetric spacetime. The
spacetime may be nonstatic, accelerating, expanding and shearing.
The integrability conditions (\ref{cons}) simplify or can be
completely solved depending on the forms of the potentials $\nu$,
$\lambda$ and $Y$. We emphasise that this is the most general
solution to the conformal Killing vector equation (\ref{lie}) for
the spherically symmetric spacetime (\ref{metric}). In the following
sections we consider some particular cases of this solution and
their relationship to exact solutions of the Einstein field
equations.

\section{Static spacetimes} \label{sec:3}

To regain the static spherically symmetric spacetimes (\ref{static})
we set
\[
\nu = \nu(r)\,, \qquad \lambda = \lambda(r)\,, \qquad Y = r
\]
Then the components of the conformal Killing vector and the
conformal factor (\ref{soln}) become
\begin{subequations}  \label{conf_static}
\begin{eqnarray}
X^0 &=& r^2 e^{-2\nu} A^i_{{}t} \eta_i + A^0  \\
X^1 &=& -r^2 e^{-2\lambda} A^i_{{}r} \eta_i + A^4
\\
X^2 &=& A^i (\eta_i)_{{}_\theta} - a_3\sin\phi + a_4\cos\phi
\\
X^3 &=& \csc^2\theta A^i (\eta_i)_{{}_\phi} - \cot\theta
(a_3\cos\phi+a_4\sin\phi) + a_6   \\
\psi &=& r^2 \eta_i \left( e^{-2\nu} A^i_{{}tt} - e^{-2\lambda}
\nu^\prime A^i_{{}r} \right) + A^0_{{}t} + \nu^\prime A^4
\end{eqnarray}
\end{subequations}
The integrability conditions (\ref{cons}) simplify to
\begin{subequations} \label{cons_static}
\begin{eqnarray}
r A^i_{{}tr} + (1-r\nu^\prime) A^i_{{}t}
&=& 0  \\
r e^{-2\nu} A^i_{{}tt} + r e^{-2\lambda} A^i_{{}rr} +
(2-r\lambda^\prime-r\nu^\prime) e^{-2\lambda} A^i_{{}r} &=& 0
\\
r^2 e^{-2\nu} A^i_{{}tt} + r(1-r\nu^\prime) e^{-2\lambda} A^i_{{}r}
+ A^i &=& 0
\\
e^{2\lambda} A^4_{{}t} - e^{2\nu} A^0_{{}r} &=& 0
\\
-A^0_{{}t} + \left( \displaystyle\frac{1}{r} -\nu^\prime \right) A^4
&=& 0
\\
-A^0_{{}t} + \left( \lambda^\prime-\nu^\prime \right) A^4 +
A^4_{{}r} &=& 0
\end{eqnarray}
\end{subequations}
where primes denote differentiation with respect to $r$.

We observe that the equations
(\ref{conf_static})-(\ref{cons_static}) are equivalent to the
corresponding system obtained by Maharaj {\em et al}.~\cite{maharaj2}.
Maartens {\em et al}.~\cite{maartens3,maartens4} demonstrated that the conformal
solution (\ref{conf_static})--(\ref{cons_static}) can be used for
categorisation and classification of static spherically symmetric
spacetimes in terms of the Weyl tensor. A number of exact solutions
to the Einstein field equations with a conformal symmetry,
corresponding to special cases of
(\ref{conf_static})--(\ref{cons_static}), have been found. These
include the solutions of Coley and Tupper \cite{coley1,coley2,coley3},
Herrera {\em et al}.~\cite{herrera1}, Herrera and Ponce de Leon \cite{herrera2},
Maartens and Maharaj \cite{maartens2}, Mak and Harko \cite{mak} and Tello--Llanos
\cite{tellollanos}.

\section{Inheriting vectors} \label{sec:4}

Herrera \textit{et al}.~\cite{herrera1} and Maartens \textit{et al}.~\cite{maartens1}
introduced the condition
\begin{equation}
{\cal L}_{\textbf{X}} u_a = \psi u_a   \label{inherit}
\end{equation}
where $\bf u$ is the fluid four--velocity. Coley and Tupper \cite{coley1}
called vectors satisfying (\ref{inherit}) inheriting conformal
Killing vectors as fluid flow lines are mapped conformally. The
existence of inheriting vectors in spherically symmetric spacetimes
has been comprehensively studied by Coley and Tupper \cite{coley2,coley3}
for particular forms of the energy momentum tensor and gravitational
potentials. Here we show that it is possible to generate the general
inheriting conformal Killing vector for the spacetime (\ref{metric})
from our conformal solution (\ref{soln})-(\ref{cons}). This solution
will be applicable to any matter distribution.

In a comoving frame of reference the fluid 4--velocity $\bf u$ is
given by $u^a = e^{-\nu} \delta^a_{{}0}$. On substitution in the
inheriting condition (\ref{inherit}) we obtain the following system:
\begin{subequations} \label{inherit_sys}
\begin{eqnarray}
\nu_t X^0 + \nu_r X^1 + X^0_{{}t} &=& \psi  \label{inherit_sys1} \\
X^0_{{}r} &=& 0  \label{inherit_sys2}  \\
X^0_{{}\theta} &=& 0  \label{inherit_sys3}  \\
X^0_{{}\phi} &=& 0  \label{inherit_sys4}
\end{eqnarray}
\end{subequations}
It is relatively straight--forward to show that the integrability
conditions (\ref{cons}) and the inheriting conditions
(\ref{inherit_sys}) may be solved in general. This essentially
follows because (\ref{inherit_sys}) implies that $X^0 = X^0(t)$ and
the integration is simplified. The components of the inheriting
conformal vector and the conformal factor are
\begin{subequations} \label{conf_inherit}
\begin{eqnarray}
X^0 &=& A^0(t)  \\
X^1 &=& A^4(r)   \\
X^2 &=& -a_3 \sin\phi + a_4 \cos\phi  \\
X^3 &=& -\cot\theta (a_3 \cos\phi + a_4 \sin\phi) + a_6  \\
\psi &=& {\dot A}^0 + \nu_t A^0(t) + \nu_r A^4(r)
\end{eqnarray}
\end{subequations}
and the integrability conditions (\ref{cons}) reduce to
\begin{subequations} \label{cons_inherit}
\begin{eqnarray}
 - \dot{A}^0 + \left( \frac{Y_t}{Y} -
\nu_t \right) A^0 + \left( \frac{Y_r}{Y} - \nu_r \right) A^4
&=& 0   \\
- \dot{A}^0 + (A^4)^{{}^\prime} + (\lambda_t - \nu_t ) A^0 +
(\lambda_r - \nu_r ) A^4 &=& 0
\end{eqnarray}
\end{subequations}
Note that the inheriting condition (\ref{inherit_sys1}) is
identically satisfied with the $X^0$, $X^1$ and $\psi$ given in
(\ref{conf_inherit}).

The integrability equations (\ref{cons_inherit}) comprise a system
of first order partial differential equations which may be solved
using the method of characteristics. The general solution is given
by
\begin{subequations} \label{soln_inherit}
\begin{eqnarray}
 \ln Y-\nu &=& F(u) + \ln A^0     \\
\lambda-\nu &=& F(u) - G(u) + \ln A^0 - \ln A^4
\end{eqnarray}
\end{subequations}
where
\[
u = \int\frac{dt}{A^0} - \int\frac{dr}{A^4}
\]
and $F$, $G$ are arbitrary functions. Thus we have generated the
general inheriting conformal symmetry (\ref{conf_inherit}) for the
spherically symmetric line element (\ref{metric}). The metric
functions $\nu$, $\lambda$ and $Y$ are restricted by
(\ref{soln_inherit}) for the existence of such a symmetry. Of course
the form of this inheriting vector will be restricted further on
application of the field equations.

It is important to note that spacetimes admitting inheriting Killing
vectors are very rare. This follows from the detailed analyses of
Coley and Tupper \cite{coley1,coley2}. They demonstrated that orthogonal
synchronous, perfect fluid spacetimes do not admit any proper
inheriting Killing vectors; the Robertson--Walker spacetimes are the
exceptional case that admit inheriting vectors. The spherically
symmetric spacetimes with known examples of inheriting vectors are
the special cases of conformal Robertson--Walker, static
Schwarzschild interior spacetimes and the generalised
Gutman--Bespalko--Wesson spacetimes. In this section we have found
the general form of the inheriting conformal symmetry for the
shearing spherically symmetric spacetime (\ref{metric}). This
generalises the results of Coley and Tupper. The generation of the
inheriting symmetry (\ref{conf_inherit}) is made possible by the
general conformal symmetry (\ref{soln})-(\ref{cons}).

\section{Exact solutions with conformal symmetry} \label{sec:5}

It is important to characterise exact solutions of the Einstein
field equations with a symmetry vector so that the geometrical
features can be studied. For the general spherically symmetric
metric (\ref{metric}), the consistency conditions (\ref{cons}) have
to be solved in conjunction with the nonlinear field equations to
generate the conformal vector $\bf X$. The nonlinearity of the
Einstein equations makes this a formidable task to achieve in
general. If we set $\nu = \nu(r)$, $\lambda = \lambda(r)$ and take
$Y(t,r)$ to be a separable function then the Einstein field
equations can be solved to admit the line element
\begin{equation}
ds^2 = -\frac{r^2}{4} dt^2 + \frac{1}{\varepsilon + cr^2} dr^2 + r^2
\left(\frac{\varepsilon}{2}+h(t)\right) (d\theta^2 + \sin^2\theta
d\phi^2) \label{exact}
\end{equation}
where
\begin{equation}
h(t) = \left\{
\begin{array}{lll}
A \sin t + B \cos t & &  \varepsilon = -1 \nonumber \\
-\frac{1}{4} t^2 + At +B & &  \varepsilon = 0 \nonumber \\
Ae^t + B e^{-t} & &  \varepsilon = 1 \nonumber
\end{array} \right.
\end{equation}
with $A$, $B$, $\varepsilon$ and $c$ being constants. A number of
authors (see Stephani \textit{et al}.~\cite{stephani}) have shown that the
expanding, accelerating and shearing spacetime (\ref{exact}) admits
the barotropic equation of state $p = \mu + 6c$. Maharaj and Maharaj
\cite{maharaj1} have shown that the class of spherically symmetric solutions
(\ref{exact}) admits a hypersurface orthogonal conformal Killing
vector ${\bf X} = (0,X^1,0,0)$ where
\begin{equation}
X^1 = \left\{
\begin{array}{lll}
D_-\, r (-1+cr^2)^{1/2} & &  \varepsilon = -1 \nonumber \\
D_0\, r^2 & &  \varepsilon = 0 \nonumber \\
D_+ \, r (1+cr^2)^{1/2} & &  \varepsilon = 1 \nonumber
\end{array} \right.
\end{equation}
and conformal factor
\begin{equation}
\psi = \left\{
\begin{array}{lll}
D_-\, (-1+cr^2)^{1/2} & &  \varepsilon = -1 \nonumber \\
D_0\, r & &  \varepsilon = 0 \nonumber \\
D_+ \, (1+cr^2)^{1/2} & &  \varepsilon = 1 \nonumber
\end{array} \right.
\end{equation}
where $D_-$, $D_0$ and $D_+$ are constants. (Note that the final
forms of $\bf X$ and $\psi$ in the Maharaj and Maharaj \cite{maharaj1} paper
contain errors which we have corrected in this solution). It is easy
to verify that the consistency conditions (\ref{cons}) are
identically satisfied with this form of $\bf X$. The special
conformal symmetry ${\bf X}= (0,X^1,0,0)$ arises because of the
simple, separable form of the metric (\ref{exact}). The existence of
this radial conformal symmetry suggests that a more detailed study
of solutions of Einstein equations, invariant under the action of
the conformal vector $\bf X$, should be pursued. In future work, we
intend to investigate the compatibility of the conformal symmetry
with homogeneity, and also with the kinematical quantities (shear,
expansion, acceleration) following the treatment of Lortan
\textit{et al}.~\cite{lmd}.

\begin{acknowledgements}
SM thanks the National Research Foundation and the University of
KwaZulu-Natal for financial support. SDM acknowledges that this work
is based upon research supported by the South African Research Chair
Initiative of the Department of Science and Technology and the
National Research Foundation.
\end{acknowledgements}

\end{document}